\documentclass[twocolumn,prb,showpacs,floats,superscriptaddress]{revtex4}
\usepackage[dvips]{graphicx}
\usepackage{amsmath,bm}
\usepackage{psfrag}
\usepackage{color}
\begin{document}

\title{Thermocapillary effects in driven dewetting and self-assembly of pulsed laser-irradiated metallic films}

\author{A.~Atena, M.~Khenner}
\affiliation{Department of Mathematics, University at Buffalo, SUNY, Buffalo, NY 14260, USA}

\newcommand{\Section}[1]{\setcounter{equation}{0} \section{#1}}
\newcommand{\rf}[1]{(\ref{#1})}
\newcommand{\beq}[1]{ \begin{equation}\label{#1} }
\newcommand{\eeq}{\end{equation} }

\begin{abstract}
In this paper the lubrication-type dynamical model is developed of a molten,
pulsed laser-irradiated metallic film.
The heat transfer problem that incorporates the absorbed heat from a single beam or
interfering beams is solved analytically.
Using this temperature field, we derive the 3D long-wave evolution PDE for the film height.
To get insights into dynamics of dewetting, we study the 2D version of the evolution equation
by means of a linear stability analysis and by numerical simulations. The stabilizing and destabilizing
effects of various system parameters, such as the peak laser beam intensity, the film optical thickness, the reflectivity,
the Biot and Marangoni numbers, etc. are elucidated.
It is observed that the film stability is promoted for such parameters variations that increase the
heat production in the film.
In the numerical simulations the impacts of different irradiation modes are investigated.
In particular, we obtain that in the interference heating mode
the spatially periodic irradiation
results in a spatially periodic film rupture with the same, or nearly equal period.
The 2D model qualitatively reproduces the results of the experimental observations
of a film stability and spatial ordering of a re-solidified nanostructures.
\end{abstract}
\pacs{47.54.r, 47.61.-k, 47.55.nb, 81.16.Dn, 81.16.Rf}

\date{\today}
\maketitle

\section{Introduction}

\label{Sec1}

Studies of dewetting, rupture and pattern formation in thin liquid films are very  important for advanced technologies
and also present an interest for basic physical sciences.
However, most experimental and theoretical results
have been obtained for aqueous and polymer films.

In this paper we report on our modeling studies of a metallic film dewetting by pulsed laser irradiation (PLI).
In the experiments on laser-irradiated metal films by Bischof \textit{et al.} \cite{BSHL},
both generic scenarios of dewetting and rupture were observed, i.e. by the growth of surface perturbations
(spinodal regime \cite{V,RJ,BWD}) and by nucleation and growth of holes \cite{GKS,SHJ}.
Very recently, several groups reported
the results of similar experiments on thinner films (thickness $h\sim 3-15$ nm) that are irradiated by
a single pulsed beam \cite{HCS,FKS,TTFSK,TFTGKS,LTGZK}, as in Bischof \textit{et al.}, as well as by two
or more interfering pulsed beams
(Pulsed Laser Interference Irradiation, PLII) \cite{KVR,FTKS,FTKKS}.
These experiments indicate that dewetting in such films is primarily spinodal.

The process of dewetting is analyzed in the cited papers using a
thermal transport modeling \cite{HCS,KVR,TTFSK,TFTGKS} and the standard isothermal thin film PDE
\cite{FKS,WD}. The consensus is that dewetting is driven
by a long-range intermolecular (van der Waals) forces and the thermocapillary forces, with negligible
material evaporation.
Most interestingly, in the PLII mode the same static interference picture
is formed on the film surface at each pulse (i.e., the alternating lines of the hot and cold regions in
two-beams PLII, or the rectangular grid of the hot and cold lines in four-beams PLII). Through the thermocapillary
fluid flow such lateral spatial
nonuniformity of the temperature field
allows the fabrication of one and two-dimensional lattices of metal nanoparticles.

While the general picture has been laid out quite clear in the cited papers, there
have been no attempts to develop a consistent PDE-based model of dewetting in a metallic film system,
which incorporates thermocapillarity and the spatial and temporal nonuniformities due to laser
irradiation. Presentation of such model is the subject of this paper.
The model
allows the studies of dewetting, rupture and  nanopatterning in 3D films
for a set of laser parameters, such as the pulse shape and repetition frequency,
the power intensity of the radiation,
the arrangement and separation distance of the
interference fringes, and
the strength of interference.

It must be noted here that the influence of
radiative heating on fluid dynamics was theoretically studied only in a handful of papers
\cite{AW,BG,H,OP,O1,G}. Most notably,
Oron \& Peles \cite{OP}  studied thermocapillary flows and instabilities of evaporative thin aqueous films with constant internal heat generation.
Oron \cite{O1} expanded this study to the case of irradiation by a single
continuous-wave laser beam.
Grigoriev \cite{G}
analyzed mechanisms of passive and active feedback control of
evaporatively driven instabilities in irradiated thin films.
The primary conclusion of these studies, which are based on long-wave theory, is that irradiation can partially suppress
the growth of instabilities.
Also, Ajaev and Willis \cite{AW} studied axisymmetric dewetting and rupture, in a molten state, of a thin metallic film,
which has been melted by
a single energetic laser pulse with a Gaussian spatial shape. They accounted for evaporation and long-range intermolecular attraction to the substrate
and identified the thermocapillary stresses as the major driving force of a film evolution.
It must be noted that neither of these papers considers nonuniform irradiation or the film reflectivity.
The latter, as has been pointed out in Refs. \cite{TTFSK,TFTGKS} is often the quantity of key importance for the dynamics of the
ultrathin metallic films.

This paper is organized as follows.
In Section \ref{Sec2} we formulate the equations of fluid motion for a film irradiated by PLI or PLII, and then
obtain the temperature field in the long-wave approximation. Using this field, we
derive the 3D long-wave evolution PDE for the surface height. In Section \ref{Sec3} we perform the
linear stability analysis of the 2D surface of the film and the numerical simulations of the 2D surface dynamics.
We show how the surface stability, surface shape, rupture time and
nanostructure array formation are
affected by key dimensionless parameters, such as the film optical thickness, peak laser beam intensity, number
of incident pulses and their duration,
Marangoni and Biot numbers, and
by the lateral spatial nonuniformity of the heat production in the PLII mode. Also we 
compare the qualitative and quantitative  
characteristics of stability and dewetting in the systems with zero and nonzero reflectivity.
Section \ref{Sec5} contains the discussion and conclusions.

\section{Derivation of the Temperature Field and Evolution Equation}

\label{Sec2}

In the PLI or PLII experiments, the solid film is  periodically melted by a laser pulse.
Following Refs. \cite{BSHL,FKS,TTFSK,TFTGKS,LTGZK,KVR,FTKS,FTKKS} we disregard periodic cycles of the melt resolidification
between laser pulses and assume, for the purpose of the analysis and nonlinear dynamical simulations of the dewetting process,
that the metallic film is always in the molten (i.e., liquid) state. This is reasonable, since it has been shown that
the mass transport in the solid state is negligible compared to the one in the liquid state,
and so is the film deformation \cite{FKS}. In other words, our model assumes that between the laser pulses the film cools
down to the temperature that is just above the solidification temperature, and thus it always stays liquid.
Correspondingly, the dewetting process is treated as a continuous one in simulations.

Thus we assume a thin film of an incompressible Newtonian liquid lying on a planar horizontal substrate.
The mean height of the film, $H$, is assumed much smaller than the lateral dimension $L$, thus
$H/L = \epsilon  \ll 1$.
The surface tension  $\tilde\sigma$ is a linear function of the temperature $\tilde T$,
\begin {equation}\tilde \sigma = \tilde \sigma_m-\tilde\gamma (\tilde T-\tilde T_m ),\label{tension} ~~~~  \tilde T >\tilde T_m.\end{equation}
Here $\tilde T_m$ is the melting temperature of the film, $\tilde \sigma_m$ is the surface tension at the melting temperature, and
$\tilde \gamma= -\frac{\partial \tilde \sigma}{\partial \tilde T} >0$. The tildes mark dimensional quantities.
The governing equations of the melt are the Navier-Stokes, continuity, and the energy equations:
\begin{equation} \rho({\bf{\tilde v}}_{\tilde t}+{\bf{(\tilde v \cdot \tilde\nabla )}\bf{\tilde v}})={\bf{\tilde\nabla \cdot \tilde \Omega }}+\rho {\bf{\tilde g}}, \label{ns}\end{equation}
\begin{equation} \tilde \nabla \cdot {\bf{\tilde v}}=0, \label{cont}\end{equation}
\begin{equation} \rho c_{p}\left(\tilde T_{\tilde t}+\tilde {\bf{v}} \cdot \tilde \nabla \tilde T\right)= \kappa{\tilde \nabla }^2\tilde T+\tilde \tau_{ij}\frac{ \partial \tilde u_i}{\partial \tilde x_j}+\tilde Q, \label{Energy}\end{equation}
where ${\bf{\tilde v}}=(\tilde u,\tilde v,\tilde w) \equiv (\tilde u_1,\tilde u_2,\tilde u_3)$ is the velocity field,
$\tilde \Omega_{ij}= -\tilde p\delta_{ij}+\tilde \tau_{ij}$ is the full stress tensor for incompressible fluid,
$\tilde \tau_{ij}=\mu \left(\frac{ \partial \tilde u_i}{\partial \tilde x_j}+\frac{ \partial \tilde u_j}{\partial \tilde x_i}\right )$ is the viscous stress tensor, $\mu$ is the dynamic viscosity, $\rho$ is the density, $\kappa$ is the thermal conductivity, ${\bf{ g}}$ is the gravitational acceleration, and $\tilde Q$ is the internal heat source.
The internal heat generation is due to the absorption of radiation from the monochromatic laser beam. We assume (i) that the
free surface of the film is
optically smooth, non-scattering, and non-emissive, and (ii) that
the solid substrate supporting the film is black (which rules out reflections from the film-substrate interface).

Under the stated assumptions
the form of the
heat source term is given by Bouguer's law \cite{Boyd}:
\begin{equation}\tilde Q = \frac{\delta I(1 -R(\tilde h))}{2}f(\tilde x, \tilde y,\tilde t)
 \exp{(\delta(\tilde z-\tilde h))}, \label{Source}\end{equation}
where $I$ is the laser power intensity, $\delta$ is the spatially uniform optical absorption coefficient, $\tilde h $ is the position of the free surface, $R(\tilde h)$ is reflectivity and
$f(\tilde x, \tilde y,\tilde t)$ is a positive function with mean value one whose
functional form depends on the interference mode and the temporal shape of the laser pulse. 

The boundary conditions  at the free surface $\tilde z=\tilde h(\tilde x, \tilde y,\tilde t)$ are:
\begin{itemize}\item[(i)] The normal and shear stress balances :
\begin{equation} {\bf{n \cdot \tilde \Omega \cdot n}}=-\tilde \sigma{\bf{\nabla \cdot n}}+\tilde \Pi, \label{normal}\end{equation}
\begin{equation}{\bf{t \cdot \tilde \Omega \cdot n}}={\bf{t \cdot \nabla }}\tilde\sigma, \quad
{\bf{n}}=\frac{\left(-\tilde h_{\tilde x},-\tilde h_{\tilde y},1\right)}{\sqrt{1+\tilde h_{\tilde x}^2+\tilde h_{\tilde y}^2}}, \label{shear}\end{equation}
where ${\bf{n}}$ is the unit outward normal to the surface, ${\bf{t}}$ is the unit tangent vector,
and
$\tilde \Pi$ is the disjoining pressure. The latter is specified in the form $\tilde \Pi=(\tilde A/6\pi)\tilde h^{-3}+\tilde B\tilde h^{-2}$ where $\tilde A$ and $\tilde B$ are the Hamaker constants. The first term is due to the dispersion (van der Waals) forces and the second term is
due to the contributions from the kinetic energy of the electrons \cite{Derjaguin}.
\item[(ii)] The kinematic condition:
\begin{equation} \tilde w=\tilde h_{\tilde t}+\tilde u\tilde h_{\tilde x}+\tilde v\tilde h_{\tilde y},\label{kinematic1}\end{equation}
that balances the normal component of the liquid velocity with the speed of the interface.
\item[(iii)] The temperature boundary condition is given by the Newton's law of cooling:
\begin{equation} \kappa\tilde T_{\tilde z}=-\alpha _h \left( \tilde T-\tilde T_a\right ), \label{tbc}\end{equation}
where $\alpha_h$ is the heat transfer coefficient and $\tilde T_a$ is the air temperature.\\
\end{itemize}

The boundary conditions for velocity of the melt flow at the solid boundary $\tilde z=0$ (the substrate) are no-slip, $\tilde u=\tilde v=0$,  and no-penetration, $\tilde w=0$.  
Two types of the temperature boundary condition at the substrate will be considered. 

The boundary condition of the first type (TBC1) is the Newton's law of cooling at $\tilde z=0$:
\begin{equation} \kappa\tilde T_{\tilde z}=\alpha_s(\tilde T- \tilde T_s),\label{tbcs1}\end{equation}
where $\alpha_s$ is the heat exchange coefficient and $\tilde T_s$ is the temperature of the substrate. Note that the
usual case of a fixed temperature at the solid boundary, $\tilde T=\tilde T_s$ (which corresponds to perfectly conducting substrate) can be easily obtained.
It suffices to take the limit $\alpha_s \rightarrow \infty$
(or equivalently, the limit $\beta_s \rightarrow \infty$, see the definition of $\beta_s$ below)
in the dimensionless expressions for the temperature and in the
evolution PDE for the film height.
It is clear from the form of these
equations that this limit exists and that taking the limit results simply in some terms dropping out of the equations.

The boundary condition of the second type (TBC2) is the continuity of the temperature and the thermal flux at $\tilde z=0$:
\begin{equation} \tilde T=\tilde \theta,\quad \kappa \tilde T_{\tilde z}= \kappa_s \tilde \theta_{\tilde z}, \label{tbcs2}\end{equation}
where $\tilde\theta$ is the temperature field in the substrate. The substrate is assumed thin, $H_s \approx H$ (where $H_s$ is 
substrate thickness). Thus the temperature field $\tilde\theta$ can be derived in the lubrication approximation.

\subsection{Nondimensionalization}

We use the following scalings to nondimensionalize the problem \cite{ODB}:
$\tilde x= x L, ~ \tilde y= y L,~ \tilde z= z H,~~\tilde h= h H,$
$\tilde u= u U, ~ \tilde v= v U,~ \tilde w= w \epsilon U$,
$ \tilde t= (L/U)t$,
$\tilde T=(IH/\kappa)T,$
$\tilde p=(\mu U/\epsilon H) p, \tilde \Pi= (\mu U/\epsilon H) \Pi, $
$\tilde \sigma =(\mu U/\epsilon)\sigma, \tilde \gamma=(\mu U\kappa/\epsilon IH)\gamma$,
where  $U$ is the characteristic flow velocity.  Typical values of the material parameters
are shown in Table I.

\subsection{Temperature Distribution in the Film}

Upon using the scalings, the dimensionless energy equation is
\begin{eqnarray}\lefteqn{\epsilon Pe \left(T_t+u T_x+ v T_y+ w T_z \right)=}&&\nonumber \\
&&T_{z z}+\epsilon^2 \left( T_{xx}+T_{yy}\right ) +(D/2)f(1-R(h))\exp{(D(z-h))}\nonumber\\
&&+\epsilon^2Br\left(u_x^2+u_y^2+v_x^2+v_y^2+w_z^2\right)\nonumber\\
&&+Br\left(u_z^2+v_z^2\right)+\epsilon^4Br\left(w_x^2+w_y^2\right),~~~~~\label {nenergy}\end{eqnarray}
where $Pe=\rho c_p U H/\kappa$ is the Peclet number, $Br=\mu U^2/HI$ is the Brinkman number, and
$D = \delta H$ is the optical thickness of a film of the uniform thickness $H$ for the incident radiation with the mean penetration length $\delta^{-1}$.
Using values of the material parameters from Table I gives
$Pe=0.019 \ll 1$ and $Br=0.11 \ll 1$. Thus in the leading order the energy equation is
\begin{equation} T_{z z}+(D/2 )f(1-R(h))\exp{(D(z-h))}=0,\label{energyapprox}\end{equation}
and the energy equation in the substrate is: 
\begin{equation} \theta_{z z}+(D/2 )f\exp{(D(z-h))}=0.\label{energyapprox_subs}\end{equation}

The dimensionless boundary conditions for Eqs. (\ref{energyapprox}) and (\ref{energyapprox_subs})  are
\begin{equation}z=h: ~~~~ T_z=-\beta (T-T_a), \label {ntbc}\end {equation}
\begin{equation} z=0: ~~~ T_z=\beta_s (T-T_s)\qquad \mbox{(TBC1)},  \label{ntbc2}\end{equation}
or
\begin{equation} \qquad \;\; T= \theta,\quad T_z= \Gamma \theta_z\qquad \mbox{(TBC2)}, \label{ntbc2a}\end{equation}
 \begin{equation} z=-h_s:~~~\theta=T_s \label{ntbc3}. \end{equation}
where $\beta=\alpha_h H/\kappa$ and $\beta_s=\alpha_s H/\kappa$ are the Biot numbers, 
$\Gamma=\kappa_s/\kappa$ is the ratio of the thermal conductivity of the substrate to one of the film,
and $h_s = H_s/H$. Of course, the boundary condition \rf{ntbc3} is not needed when the boundary condition
(TBC1) is used.

Solution of Eq. (\ref{energyapprox}) subject to the boundary conditions  (\ref{ntbc}) and 
(\ref{ntbc2}), or the solution of Eqs. (\ref{energyapprox}) and (\ref{energyapprox_subs})
subject to the boundary conditions (\ref{ntbc}), (\ref{ntbc2a}), and (\ref{ntbc3})
gives the temperature field in the film,
\begin{eqnarray}T(x,y,z,t)&=& \left(\frac{\Upsilon}{2}-\exp{(-D h)}\left(\frac{\Upsilon}{2}+ K\right)+\right.\nonumber\\
&&\left.K\exp{(D(z-h))}+\frac{z}{2}\right)f(1-R(h))+T_s+\nonumber\\
&&\left(T_a-T_s+F(h)f(1-R(h))\right)\left(\Upsilon+z\right)\beta,\nonumber\\ ~~\label{expTRefl}
\end{eqnarray}
where $K=-1/2D$,
\begin{equation}
F(h) = -\frac{\Upsilon}{2}+\exp{(-D h)}\left( \frac{\Upsilon}{2}+K\right)-\frac{h}{2}-K,
\end{equation}
and $\Upsilon = 1/\beta_s$ (TBC1), or $\Upsilon = h_s/\Gamma$ (TBC2). Note that Eq. \rf{expTRefl} results upon 
the linearization in $\beta$ of the full solution of the problem. This is warranted since $\beta \ll 1$ (see Table II).

\noindent
Substitution $z=h$ in Eq. (\ref{expTRefl}) gives the temperature at the free surface:
\begin{eqnarray} T^{(h)} &\equiv& T(x,y,h,t) = T_s -F(h)f(1-R(h))+ \nonumber \\
&& \left (\Upsilon+h\right)(F(h)f(1-R(h))+T_a-T_s)\beta.\nonumber\\ \label{surfaceTRefl}
\end{eqnarray}

\begin{figure}[!t]
\centering
		\includegraphics[width=3.4in,height=3.0in]{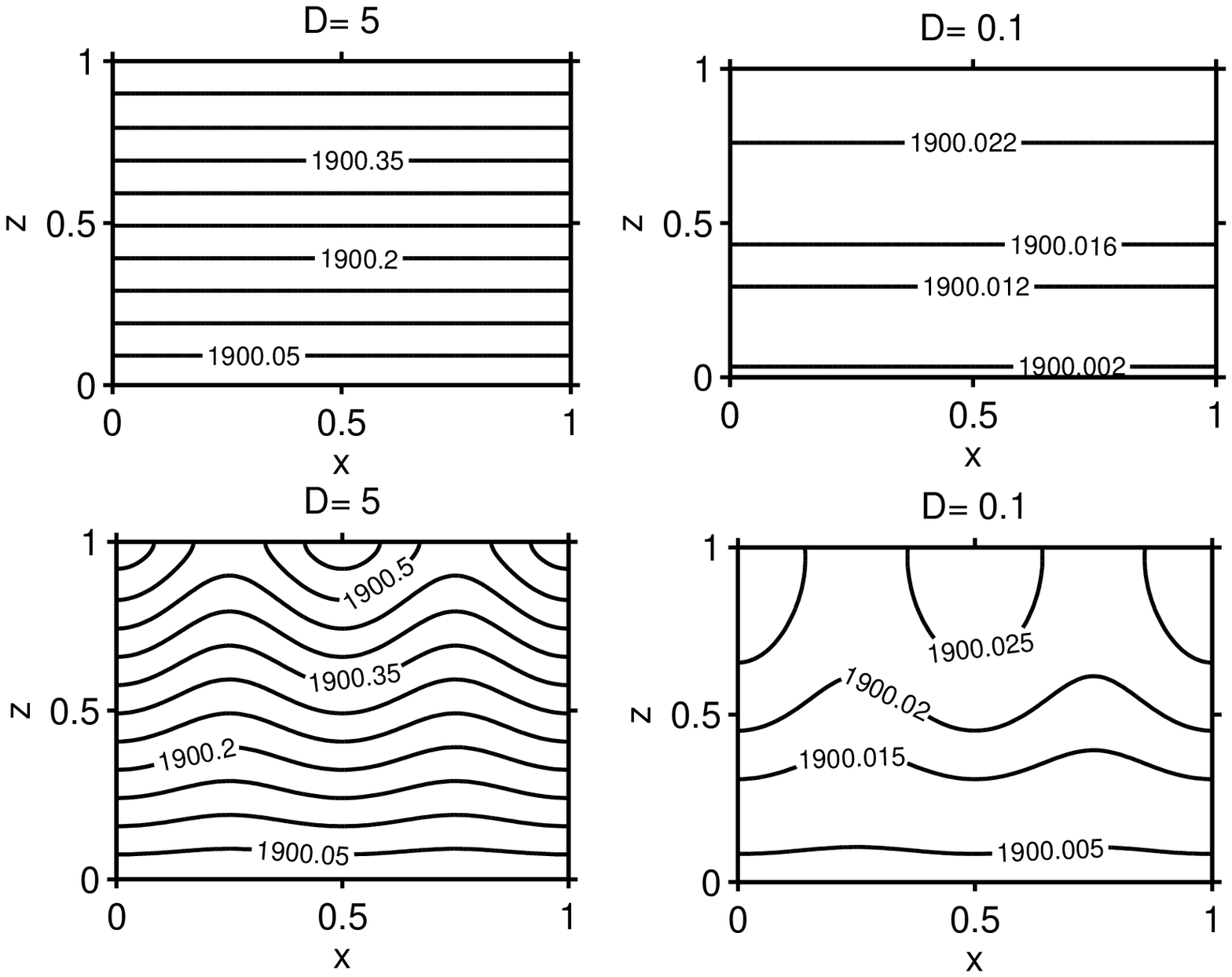}
		\caption{{\small{Contour plot of the temperature field in the film at $y=0.5$ for the case (TBC1) and $R(h)=0$. Top row: the surface is heated by the uniform (spatially and temporally) laser beam. Bottom row: the surface is heated by the temporally uniform but spatially nonuniform laser beam. Left panel: optically thick film. Right panel: optically thin film.   Notice very small
vertical temperature gradient in all cases ($\partial T/\partial z > 0$).}}}
	\label{Temp}
\end{figure}

\begin{figure}[!t]
\centering
		\includegraphics[width=3.4in,height=3.0in]{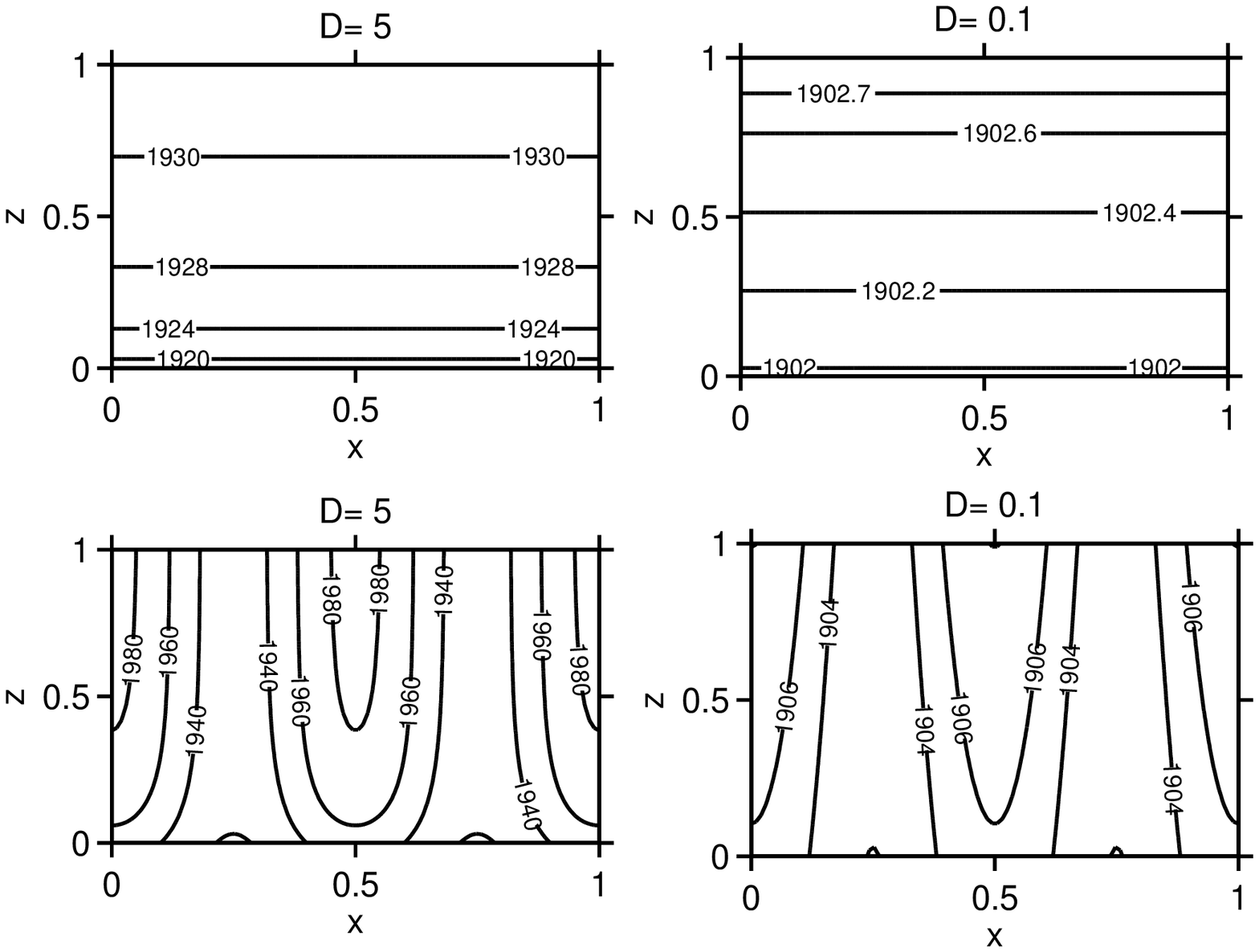}
		\caption{{\small{Same as Fig. \ref{Temp}, but for the case (TBC2) and nonzero $R(h)$ given by Eq. \rf{Rofh}. Note that vertical temperature gradient (positive) is larger than in Fig. \ref{Temp}.}}}
	\label{TempRefl}
\end{figure}

In Figs. \ref{Temp} and \ref{TempRefl} we show the typical contour plots of the temperature field in the film of the uniform dimensionless height $h=1$ at a vertical cross-section in the middle of the domain. 
These figures were obtained with (TBC1), $R(h)=0$ and (TBC2), $R(h)\neq 0$, respectively. 
(\textit{Remarks:} The form of the reflectivity function $R(h)$ is shown in Eq. \rf{Rofh} below. Unless noted otherwise in the text or in a figure caption, the parameter values that are used to obtain all figures in the paper are taken from Table II. Also, the dimensionless quantities are plotted in all Figures, except Figs. \ref{MaxT_bRefl} and
\ref{NP}.)
In the top row of both Figures, the film surface is heated by the
spatially and temporally uniform laser beam ($f=1$), while in the bottom row it is heated by the laser beam with the
uniform temporal intensity distribution but nonuniform
spatial intensity distribution
$f(x,y)=1+0.1(cos(4\pi(x-0.5))+cos(4\pi(y-0.5)))$, corresponding to four-beams PLII. If the film is  irradiated uniformly, the temperature remains uniform in any horizontal plane in the film (top rows of the Figures), but for spatially nonuniform irradiation the temperature in the film follows the shape of the intensity distribution as shown in the bottom rows of the Figures.

If the optical thickness $D<<1$ the radiation passes through the film and such films are called optically thin or transparent. If $D>>1$ the radiation penetrates only into a very thin boundary layer adjacent to the free surface of the film and in this case the film is called optically thick or opaque.  Note that in the dimensionless energy equation, Eq. (\ref{energyapprox}), $f$ characterizes the variation of the heat source in a horizontal plane, whereas the
function $a(z)=\exp{(D(z-h))}$ describes its dependence on the depth of the film, $z$. Since $0 \leq z \leq h$, $a(z)$ is the increasing function of the height $z$, however for the optically thin film the difference $a(h)-a(0)$ is a small quantity. In other words, the top and the bottom of the film receive approximately same energy from the laser beam. As a result the temperature difference across the film is small (see the right panels of Figs. \ref{Temp} and \ref{TempRefl}). On the other hand, the bottom part of the optically thick film receives less energy from the laser beam than the top part, which makes the temperature difference across the film larger (see the left panels of these Figures).


Before we proceed further, we state the form of the effective film reflectivity that we use in the paper.
We assume the following form after Refs. \cite{TTFSK,TFTGKS}:
\begin{equation}
R(h)=r_0(1-\exp(-a_r h)),
\label{Rofh}
\end{equation}
where $r_0$ and $a_r$ are the material dependent parameters, see Tables I and II.

\subsection{Derivation of the Evolution Equation for the Film Height}

In this section we outline the conventional procedure \cite{ODB} of the derivation of the evolution equation.

The dimensionless Navier-Stokes and continuity equations read:
\begin{eqnarray}\epsilon Re(u_t+uu_x+vu_y+wu_z)&=&-p_x+\nonumber\\
&&\epsilon^2\left(u_{xx}+u_{yy}\right)+u_{zz},\nonumber\\
 \hspace{-0.5 in}\epsilon Re(v_t+uv_x+vv_y+wv_z)&=&-p_y+\nonumber\\
&&\epsilon^2\left(v_{xx}+v_{yy}\right)+v_{zz},\nonumber\\
 \hspace{-0.1 in}~\epsilon^3 Re(w_t+uw_x+vw_y+ww_z)&=&-p_z+\nonumber\\
&&\epsilon^4\left(w_{xx}+w_{yy}\right)+\nonumber\\
     &&\epsilon^2 w_{zz}-G,\label{nns3}\\
 u_ { x}+ v_ { y}+ w_ { z}&=&0,\label{ncont}\end{eqnarray}
 where $Re=\rho U H/\mu$ is the Reynolds number (which is of order one in magnitude for the case considered here), and $G$  is the gravity number. In the leading order the Navier-Stokes and continuity equations read
 \begin{eqnarray}p_x&=&u_{zz}, ~~~~p_y=v_{zz},~~~p_z=-G\label{ns3approx},\\
 &&u_ {x}+ v_ {y}+ w_ {z}=0. \label{contapprox}\end{eqnarray}
Using the long-wave approximation, $|h_x|$, $|h_y| <<1$ and the usual assumption of large surface tension \cite{ODB},
$\hat C= \epsilon^{3} C^{-1}$ (where $C$ is the capillary number)
in Eq. (\ref{normal}), the dimensionless normal  stress balance condition is
 \begin{equation} -p=\hat C \left( h_{xx}+h_{yy}\right)+\frac{A}{h^3}+\frac{B}{h^2}, \label{nnormal}\end{equation}
 where  $A$ and $B$ are the dimensionless Hamaker constants. The dimensionless shear stress balance condition reads
 \begin{equation}u_z=\sigma_x+h_x\sigma_z ,~~v_z=\sigma_y+h_y\sigma_z, \label{nshear}\end{equation}
where the dimensionless surface tension is given by $\sigma= \sigma_m-\gamma (T-T_m)$.

Taking the cross-sectional averages of  $u$ and $v$  over the film height and integrating Eq. (\ref{contapprox}) using the no-slip condition, we obtain a more convenient form of the kinematic boundary condition:
 \begin{equation}h_{ t}+\left(h\overline{u}\right) _{ x}+\left(h\overline {v}\right)_{ y}=0.\label{kinematic}\end{equation}
%
Integrating the first two equations in Eq. (\ref{ns3approx}) twice in $z$ and applying the no-slip boundary condition yields
\begin{eqnarray} u&=&(\frac{z^2}{2}-hz)p_x+z u_z|_{z=h},\nonumber\\
v&=&(\frac{z^2}{2}-hz)p_y+z v_z|_{z=h}. \label{uv}\end{eqnarray}
Cross-sectional averaging of Eqs. \rf{uv} results in
\begin{eqnarray}\overline{u}&=&\frac{1}{h}\int_{0}^{h} u dz=-\frac{h^2}{3}p_x+\frac{h}{2}u_z|_{z=h},\nonumber\\
\overline{v}&=&\frac{1}{h}\int_{0}^{h} v dz=-\frac{h^2}{3}p_y+\frac{h}{2}v_z|_{z=h}.\label{uvavg}\end{eqnarray}
Since $u_z$ and $v_z$ are expressed in terms of the derivatives of the the surface tension, Eq. (\ref{nshear}), which in turn is expressed in terms of the derivatives of the temperature at the free surface, we need to calculate the latter derivatives. Hence, by differentiating Eq. (\ref{expTRefl}) and evaluating the derivatives at $z=h$ we find
\begin{eqnarray} T^{(h)} _{x(y)}&=&f(1-R(h))\left(1-\beta(h+\Upsilon)\right)F_1(h) h_{x(y)}\nonumber\\
&&fR^{'}(h)\left(1-\beta(h+\Upsilon)\right)F(h) h_{x(y)}\nonumber\\&&-F(h)(1-R(h))f_{x(y)}
\\&&+(h+\Upsilon)F(h)(1-R(h))f_{x(y)}\beta, \nonumber\\
T^{(h)}_z&=&\beta\left(f(1-R(h))F(h)+T_a-T_s\right), \label{TxTyTz} \end{eqnarray}
where 
\begin{equation} F_1(h)=\frac{1}{2}+\frac{1}{2}(\Upsilon D-1)\exp{(-D h)}. \label{F1}\end{equation}
Substituting Eq. (\ref{TxTyTz}) in Eq. (\ref{nshear}) through the derivatives of $\sigma$ and using Eq. \rf{nnormal} in Eqs. \rf{uvavg}, the expressions for the average velocities become:
\begin{eqnarray} \overline{u}&=&-\frac{Gh^2}{3}h_x+\frac{\hat C h^2}{3}(h_{xxx}+h_{yyx})+\nonumber \\
&&(\frac{A}{h^2}-\frac{2B}{3h}-Mh T^{(h)}_z)h_x-MhT^{(h)}_x+\textit{O}(\beta f_x) , \nonumber\\
 \overline{v}&=&-\frac{Gh^2}{3}h_y+\frac{\hat C h^2}{3}(h_{xxy}+h_{yyy})+\nonumber\\
 &&(\frac{A}{h^2}-\frac{2B}{3h}-Mh T^{(h)}_z)h_y-MhT^{(h)}_y+\textit{O}(\beta f_y),\nonumber\\ \label{uvavg2}\end{eqnarray}
where $M=\gamma/2$ is the Marangoni number.

Finally, substituting Eqs. (\ref{uvavg2}) in the kinematic boundary condition Eq. (\ref{kinematic})
and using Eqs. (\ref{TxTyTz})
results in the evolution equation for the film height:
\begin{eqnarray}
 h_t  & =& \nabla \cdot \left[ \frac{-\hat C}{3}h^3\nabla \nabla^2h+\frac{G}{3}h^3\nabla h-\left(\frac{A}{h}-\frac{2B}{3}\right)\nabla  h\right.\nonumber\\
 &&+M\beta (T_a-T_s)h^2\nabla h
 +MF_1(h)f(1-R(h))h^2\nabla h\nonumber\\
 &&+MR'(h)F(h)fh^2\nabla h\nonumber\\
 &&-M\beta(h+\Upsilon)R'(h)F(h)fh^2\nabla h\nonumber\\
 &&+\left.M\beta f(1-R(h))\left(F(h)-(h+\Upsilon)F_1(h))\right)h^2\nabla h\right].\nonumber \\\label{evohRef}
 \end{eqnarray}
Here the prime denotes differentiation.

In the following sections of the paper, we assume the laser irradiation either uniform in the $xy$-plane,
or nonuniform only in the $x$-direction. The former situation corresponds to the PLI case, i.e. one of a
single incident laser beam, where the interference is absent. (If there is only one beam, we approximate its spatial
intensity distribution on irradiated domain by a constant, i.e. $f=1$ or $f=f(t)$.)
The latter situation corresponds to two-beams PLII.
In both cases the simpler 2D model provides valuable insight into the dynamics of dewetting.
This model is studied in Section \ref{Sec3}.

It must be noted that in Eq. \rf{evohRef} we omitted the term $-M(1-R(h))\nabla \cdot \left[F(h)h^2\nabla f\right]$ and the similar term proportional
to the small parameter $\beta$. 
In PLII, when $f$ is nonuniform in the plane (see Eq. \rf{fc}), the omitted terms describe the surface shape change by thermocapillarity due to
in-plane temperature equilibration by heat conduction. In the PLII experiments the in-plane heat fluxes are negligible
because each pulse last only a few nanoseconds, heat losses to the substrate are large and the distance in the plane
between the interference fringes is much larger than the film thickness. Thus the lateral temperature profile
is approximately static, i.e. it is determined by the geometrical arrangement of the interference fringes.
A few sample computations that we performed with Eq. (\ref{evohRef}) where the omitted terms are present, show that these terms are
indeed much smaller than the other terms, and their influence on the film dynamics is negligible.
This can be also understood by noticing that in the experiment (and correspondingly, in our modeling) the wavenumber of the surface perturbation, $k$, is larger or much
larger than the wavenumber, $q$, of the spatially modulated laser light field $f$. See, for example,
Fig. \ref{twobeam} which is obtained with $k=2.2$ and $q=0.157$. Thus $\nabla f \sim qf \ll \nabla h \sim kh$, and
the terms proportional to $\nabla f$ are smaller than the terms proportional to $\nabla h$.

\begin{table}
 \begin{tabular}{l l}
\hline\hline
Physical Parameter & Typical values  \\
\hline
 Film thickness ($H$)&						$10~ nm$\\
 Optical absorption coefficient ($\delta$)&$10^{8} ~ m^{-1}$\\
 Film density ($\rho$ )&				$8.92*10^3 ~ Kg/m^3$\\
 Heat capacity ($c_p$)&					$420 ~ J/Kg K$\\
 Thermal conductivity (film) ($\kappa$)&	$100 ~ W/m K$\\
 Thermal diffusivity ($\chi $)&	$2.675*10^{-5}~ m^2/s$\\
 Melting temperature ($\tilde T_m$) of Co  	&	$1768 ~ K$\\
 Ambient Temperature ($\tilde T_a$)&$300 ~ K$\\
 Substrate Temperature ($\tilde T_s$)&$1900 ~ K$\\
 Acceleration of gravity ($\tilde g$)&$9.8~  m/s^2$\\
 Dynamic Viscosity ($\mu$)&	$4.45*10^{-3}~ Pa-s$\\
 Surface tension (at melting pt.)($\tilde \sigma_m$)&$1.88 ~J/m^2$\\
 $\tilde\gamma$ (at melting point) &$5*10^{-4} ~J/K m^2$\\
Characteristic velocity ($U=\tilde \mu/\rho H$)&	$50~ m/s$\\
 Peak Intensity ($I$) 		&							$10^{10}~ W/m^2$\\
Heat transfer coefficient ($\alpha_h$)&$1.41*10^4~ W/m^2K$\\
Heat transfer coefficient ($\alpha_s$)& $>>1~ W/m^2K$\\
Hamaker constant ($\tilde A$)&$1.41*10^{-18}~J$\\
Hamaker constant ($\tilde B$)&$2.6*10^{-13}~N$\\
Substrate thickness ($H_s$)& $10~ nm$\\
Thermal conductivity (substrate) ($\kappa_s$)&	$1.3 ~ W/m K$\\
Inverse exponent,\\ reflectivity function ($a_r^{-1}$)&$15.5~ nm$\\
\hline\hline
\end{tabular}
\caption{{\small{Material parameters. 
}}}
\end{table}
\begin{table}
\begin{tabular}{lll}
\hline\hline
Dimensionless group&Definition&Typical values \\
\hline
Scaling parameter ($\epsilon$)&$H/L$&0.01\\
Reynolds number ($Re$)&$\rho U H/\mu$&1\\
Brinkman number ($Br$)&$\mu U^2/HI$&$0.11$\\
Peclet number ($Pe$)&$\rho c_p U H/\kappa$&0.019\\
Capillary number ($C$)&$\mu U/\sigma$&0.1184\\
Gravity parameter ($G$)&$\epsilon \rho g H^2/\mu U$&$3.93*10^{-13}$\\
Biot number ($\beta$)&$\alpha_h H/\kappa$ &$10^{-6}$\\
Biot number ($\beta_s$)&$\alpha_s H/\kappa$ &$10^2-\infty$\\
Surface tension ($\sigma_m$)&$\epsilon \tilde\sigma_m/\mu U$&0.084\\
Marangoni number ($M$) &$\epsilon I H\tilde\gamma/2\mu U \kappa$&$1.125*10^{-5}$\\
Hamaker constant ($A$)&$\epsilon \tilde A/6 \pi\mu U H^2$&$3.37*10^{-5}$\\
Hamaker constant ($B$)&$\epsilon \tilde B/\mu U H $&$1.17*10^{-6}$\\
Melting temperature ($T_m$)&$\kappa \tilde T_m/I H$ &1768\\
Ambient temperature ($T_a$)& $\kappa \tilde T_a/I H$ &300\\
Substrate temperature ($T_s$)&$\kappa \tilde T_s/I H$  &1900\\
Optical thickness ($D$) &$\delta H$ &$1$\\
Substrate thickness ($h_s$) & $H_s/H$ & $1$\\
Ratio of thermal\\conductivities ($\Gamma$)&$\kappa_s/\kappa$&$1.3*10^{-2}$\\
Pre-factor,\\ reflectivity function &$r_0$& $0.44$\\
\hline\hline
\end{tabular}
\caption{{\small{Dimensionless parameters. 
}}}
\end{table}

\section{The 2D Model of film dynamics}

\label{Sec3}

The 2D reduction of the evolution equation \rf{evohRef} reads:
\begin{eqnarray}
h_t&=&\left[-\frac{\hat C}{3}{h}^3h_{xxx}+\frac{G}{3}{h}^3h_x-\left( \frac{A}{h}-\frac{2B}{3}\right )h_x \right.\nonumber \\
&&+M\beta(T_a-T_s){h}^2h_x+MF_1(h)f(1-R(h)){h}^2h_x\nonumber\\
&&+MR'(h)F(h)fh^2h_x
-M\beta(h+\Upsilon)R'(h)F(h)fh^2h_x\nonumber\\
&&+M\beta f(1-R(h))\left(F(h)-(h+\Upsilon)F_1(h)\right)h^2h_x
]_x.\nonumber \\
\label{E0Refl}\end{eqnarray}

\subsection{Linear Stability Analysis}

\label{Sec3a}

For the purpose of the linear stability analysis, we assume uniform laser power intensity distribution ($f=1$) and a
small normal perturbation of the uniform base state, $h=h^{(0)}+\xi(x,t)=1+e^{\omega t}e^{ikx}$, where $h^{(0)}$ 
equals to one due to nondimensionalization, and $\omega$ represents the growth rate of the perturbation having a wave number $k$. Linearizing Eq. (\ref{E0Refl}) in $\xi$ results in the dispersion relation:
\begin{eqnarray} \omega(k)&=&-\frac{G}{3}k^2-\frac{\hat C}{3}k^4+(A-\frac{2B}{3})k^2-M\beta(T_a-T_s)k^2\nonumber\\
&&+MR'(1)F(1)(-1+\beta(1+\Upsilon))k^2\nonumber\\
&&\hspace{-0.5 in}+M(1-R(1))\left(-F_1(1)-\beta\left(F(1)-(1+\Upsilon)F_1(1)\right)\right)k^2.\nonumber\\\label{omegaRefl}\end{eqnarray}
From Eq. \rf{omegaRefl} we find the critical wave number $k_c$ such that $\omega>0$ for $0<k<k_c$:
\begin{eqnarray}k_c &=&\left[\frac{3}{\hat C}\left(-\frac{G}{3}+A-\frac{2B}{3}-M\beta(T_a-T_s)\right.\right.\nonumber\\
&&+MR'(1)F(1)(-1+\beta(1+\Upsilon))\nonumber\\
&&\hspace{-0.5 in}\left.\left.+M(1-R(1))\left(-F_1(1)-\beta\left(F(1)-(1+\Upsilon)F_1(1)\right)\right)\right)\right]^{1/2}\nonumber \\\label{cutoffRefl}\end{eqnarray}
The growth rate $\omega$ attains its maximum value $\omega_{max}$ at the wave number $k_m=k_c/\sqrt{2}$, which is usually referred to as the "most dangerous mode".

The terms at the right-hand-side of Eq. (\ref{omegaRefl}) describe:
the stabilizing effect of gravity, the stabilizing effect of capillary forces, the destabilizing and stabilizing effects of the van der Waals component and the electronic component of the disjoining pressure, respectively, the stabilizing effect of the temperature gradient across the film, if $T_a>T_s$, and destabilizing effect otherwise, and the last two terms are  due to  the volumetric heat source. The coefficient of the last two terms in Eqs. (\ref{omegaRefl}) and (\ref{cutoffRefl}), $MR'(1)F(1)(-1+\beta(1+\Upsilon))+M(1-R(1))\left(-F_1(1)-\beta\left(F(1)-(1+\Upsilon)F_1(1)\right)\right)$  is negative for the typical parameters values from Table II and for all values of $D$. 
This holds true for both cases of the boundary conditions ((TBC1) or (TBC2)) and does not depend on the presence of 
film reflectivity (see Figs. \ref{KcVdH} and \ref{coeffRefl}).
Thus the term associated with the heat source has a stabilizing impact \cite{OP,O1,G}.  


\subsubsection{Results for the case (TBC1) and $R(h)=0$}

In this section $\Upsilon = 1/\beta_s$ and $r_0=0$ in Eq. \rf{Rofh}.

\begin{figure}[!h]
\centering
	\includegraphics[width=3.4in,height=2.5in]{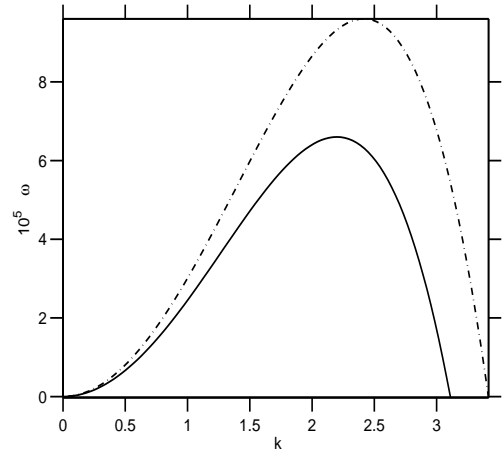}
		\caption{{\small{Variation of $\omega$ with $k$: The dash-dot curve shows $\omega(k)$ calculated without the term containing the effect of the heat source  in Eq. (\ref{omegaRefl}), the solid curve shows $\omega(k)$ calculated with all terms included. }}}
	\label{omegaVk}
\end{figure}

The typical graphs of $\omega(k)$ are shown in Fig. \ref{omegaVk}. The
solid curve represents $\omega(k)$ calculated with all terms at the right-hand-side of Eq. (\ref{omegaRefl}) and the
dash-dot curve shows $\omega(k)$ computed without the last term. (The fifth term is automatically zero since $R'(1)=0$
due to $R(h)=0$.) As expected both the maximum growth rate and the cutoff wave number in the former case are smaller than the corresponding quantities in the latter case.

\begin{figure}[!h]
\centering
	\includegraphics[width=3.4in,height=2.2in]{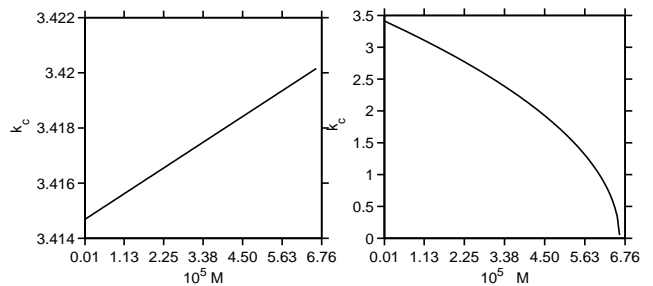}
	\caption{{\small{Variation of $k_c$ with $M$: The left panel is the graph of $k_c$ calculated without the term containing the effect of the heat source and the right panel is the graph of $k_c$ calculated with all terms included. }}}
	\label{KcVM}
\end{figure}

The Marangoni effect is expressed by the two components, one responsible for the destabilizing thermocapillary effect
(the fourth term in Eq. (\ref{omegaRefl}) with $T_a < T_s$, see Table II), and another responsible for the stabilizing effect associated with the heat
source (the fifth term). Fig. \ref{KcVM} shows the Marangoni effect when the stabilizing component is absent (left panel) and when both components are present (right panel). Since the Marangoni number $M$ increases with the peak intensity $I$,  the irradiation of the film with a high intensity laser has stabilizing influence (below the onset of evaporation).

\begin{figure}[!h]
\centering
	\includegraphics[width=3.4in,height=2.5in]{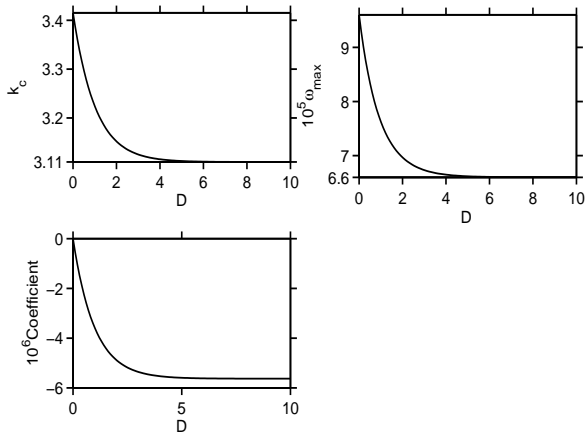}
		\caption{{\small{Variation of $k_c$ (left), variation of $\omega_{max}$ (right), and variation of the coefficient $M\left(-F_1(1)-\beta\left(F(1)-(1+\Upsilon)F_1(1)\right)\right)$ (bottom) of the last term in Eqs. (\ref{omegaRefl}) and (\ref{cutoffRefl})  with optical thickness $D$. Here $\Upsilon = 1/\beta_s$ (case (TBC1)).
}}}
	\label{KcVdH}
\end{figure}

The impact of the optical thickness $D$ on the stability of the perturbation is shown in
Fig. \ref{KcVdH}.
It can be seen that $k_c$ and $\omega_{max}$  both decrease as $D$ increases, while the magnitude of the 
stabilizing effect (i.e., the coefficient) increases with $D$.
Thus in agreement with Refs. \cite{OP,O1,G}, as the film becomes more opaque the internal heat generation
increases and the associated Marangoni effect stabilizes the film.

Finally, we discuss impacts of $\beta_s$ and $\beta$. The cutoff wave number increases as $\beta_s$ increases ( $k_c\approx 3.1099$ for $\beta_s=1$ and $k_c\approx 3.2257$ for $\beta_s=10^{4}$)  approaching the constant value $\approx 3.23$ as 
$\beta_s \rightarrow \infty$. The maximum growth rate is very insignificantly affected by the change of $\beta_s$,
increasing slowly and approaching value $7.65 \times 10^{-5}$ as $\beta_s \rightarrow \infty$.
Thus the film is more stable for smaller $\beta_s$. 
Similarly, the cutoff wave number $k_c$ and the maximum growth rate increase insignificantly with increasing 
$\beta$ (while $\beta$ is kept typically small, as required by Table II). 
Clearly, as $\beta$ or $\beta_s$ increase the amount of heat in the film decreases due to heat losses to the ambient, and the film becomes less stable.

\subsubsection{Results for the case (TBC2) and $R(h)\neq 0$}

\label{Sec4}

In this section $\Upsilon = h_s/\Gamma$ and $r_0=0.44$ in Eq. \rf{Rofh}.

\begin{figure}[!h]
\centering
	\includegraphics[width=3.4in,height=2.5in]{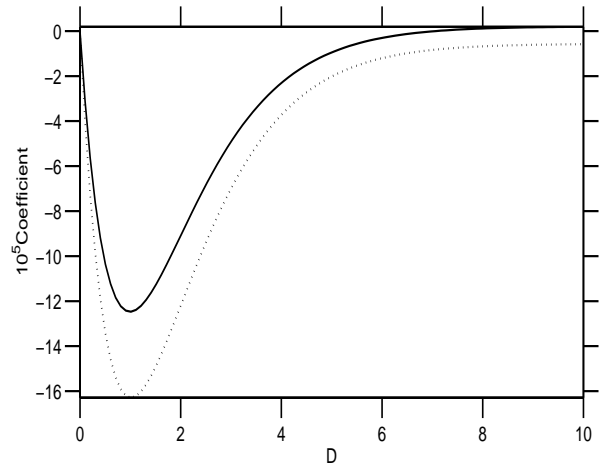}
		\caption{{\small{Variation of the coefficient, $MR'(1)F(1)(-1+\beta(1+\Upsilon))+M(1-R(1))\left(-F_1(1)-\beta\left(F(1)-(1+\Upsilon)F_1(1)\right)\right)$, of the last two terms in Eqs. \rf{omegaRefl} and \rf{cutoffRefl} with $D$. Dot curve: $R(h)=0$; solid curve: $R(h)\neq 0$. Here $\Upsilon = h_s/\Gamma$ (case (TBC2)). The dot curve is included for comparison.}}}
	\label{coeffRefl}
\end{figure}

The coefficient of the two stabilizing terms in the growth rate equation \rf{omegaRefl} is plotted in Fig. \ref{coeffRefl}.
Unlike Fig. \ref{KcVdH}, the dependence of the coefficient on $D$ is non-monotone for both zero and non-zero reflectivity when the heat 
conduction in the thin substrate is taken into account. 
The maximum stabilization occurs in films with $D \sim 1$, and the minimum one in films with $D \ll 1$
or $D\gg 1$. 

\begin{figure}[!h]
\centering
	\includegraphics[width=3.4in,height=2.5in]{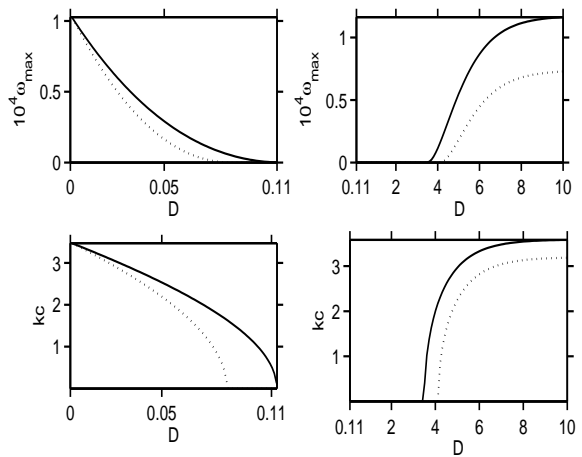}
		\caption{{\small{Variation of $\omega_{max}$ (top row) and $k_c$ (bottom row) with $D$. Dot curve: $R(h)=0$; solid curve: $R(h)\neq 0$. The dot curve is included for comparison.}}}
	\label{omaxRefl}
\end{figure}

The corresponding maximum growth rate and the cutoff wavenumber are shown in Fig. \ref{omaxRefl}.
As $D$ starts to increase from zero, the film becomes more stable. In the interval $0.11\le D\le 3.45$ complete
stabilization occurs, i.e. the growth rate is negative for all wavenumbers. For $D$ larger than 3.45, the growth rate
is positive for $0<k<k_c$ (where $k_c$ is shown in the bottom right panel), and as $D$ increases the film
becomes less stable. Such non-monotonous dependence of stability characteristics on the optical thickness
is, in our opinion, very interesting and unexpected. \textit{When the heat 
conduction in the thin substrate is taken into account, the film heated uniformly
can be completely stabilized against small perturbations in some interval of the optical thickness parameter.}
Of course, the film is still unstable for perturbations with amplitudes larger than some critical one.

\begin{figure}[!h]
\centering
		\includegraphics[width=3.4in,height=2.2in]{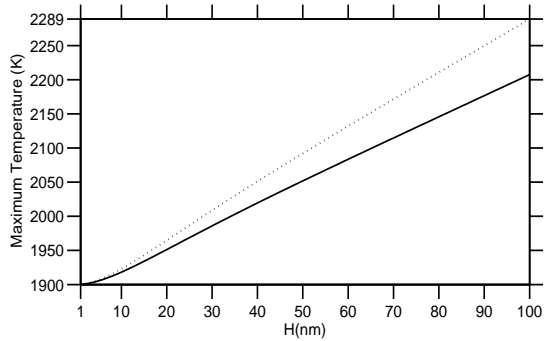}
		\caption{{\small{Plot of the maximum dimensional film temperature vs. film height. Dot curve: $R(h)=0$; solid curve: $R(h)\neq 0$. The dot curve is included for comparison.}}}
	\label{MaxT_bRefl}
\end{figure}

Fig. \ref{MaxT_bRefl} shows the \textit{maximum} temperature in the film as a function of the dimensional film height, for $f=1$.
(Of course, the dimensionless parameters were re-calculated for each new value of $H$.)
The temperature is increasing as the height inreases, confirming the observations in
Refs. \cite{BSHL,HCS,FKS,TTFSK,LTGZK,KVR,FTKS,FTKKS}.
The trend in Fig. \ref{MaxT_bRefl} also signals that higher laser power intensity is required to melt thinner films
\cite{BSHL,HCS,FKS,TTFSK,LTGZK,KVR,FTKS,FTKKS}.
The slope of the line is larger in the case of $R(h)=0$, which shows that neglecting the reflectivity over-predicts the temperature in the film. This effect is also discussed in Ref. \cite{TTFSK}.

It can be easily shown that the total heat generated in the film, i.e, the integral of the source term  $Q$ (Eq. (\ref{Source})) over the film height $h$, is an increasing function of $h$ when $\delta$ is greater than $a_r$ (see also Refs. \cite{TTFSK,TFTGKS}).  
  This is the reason
why the film temperature increases as the film thickness increases.

\subsection{Nonlinear Evolution of the Film}

The 2D  evolution equation for the film height, Eq. (\ref{E0Refl}),
is solved numerically using the method of lines. Integration in time is performed using the ODE solver RADAU whereas the discretization in space is carried out in the conservative form using the positivity-preserving \cite{ZB}, second order-accurate finite differencing on a spatially uniform staggered  grid.

The perturbation
\begin{equation}h(x,0)= 1+A_{0}cos(k_mx),\label{IC}\end{equation} where $A_{0}$ is the small amplitude and $k_m$ is the wave number of the fastest growing instability as identified in the linear stability analysis, is imposed as the initial condition. Eq. (\ref{E0Refl}) is solved  in the spatial domain $0\le x\le 2n\pi/k_m$ (where $n$ is an integer) subject to the periodic boundary conditions.

We performed simulations using different forms of $f(x,t)$ that characterizes the spatio-temporal power intensity distribution of the PLI or PLII at the film surface.

Firstly, we assume $f$ uniform in both time and space, i.e. $f\equiv 1$. This corresponds to a steady heating of the film surface by a single laser beam with a uniform spatio-temporal shape. The results of the nonlinear simulations in this regime
can be compared to the linear stability analysis in the previous section.

Next, with $f$ still spatially uniform we assign a Gaussian temporal shape to $f$, so that
\begin{equation}f\equiv f(t)=e^{-\frac{(t-T/2)^2}{2\sigma^2}} \label{sp}\end{equation} for a single pulse laser irradiation, and \begin{equation}f\equiv f(t)=\sum^{N-1}_{k=0} e^{-\frac{\left(t-(2k+1)T/2\right)^2}{2\sigma^2}} \label{np}\end{equation} for a sequence of $N$ pulses of irradiation, where $T$ is the pulse duration and  $\sigma$ is the standard deviation. (Note that $d=2\sigma\sqrt{2\ln {2}}$ is the Gaussian full width at half-maximum.) Both situations again correspond to heating of the film surface by a single laser beam.
	
Finally, we consider the case of the two-beam interference. The two-beam interference produces a spatially modulated light field having the form \cite{KVR}
\begin{equation}
f\equiv f(x)=1+\alpha \cos(q(x-\frac{\pi}{k_m})),
\label{fc}
\end{equation}
where the parameter $0<\alpha < 1$
models the strength of the interference and $2\pi/q=\ell$ is the distance between two neighboring interference fringes. This distance $\ell$ is given by $\ell=\lambda/2\sin(\theta/2)$,  where $\lambda$ is the wavelength of the primary laser beam and $\theta$ is the angle between the two interfering beams. Note that the subtraction of $\pi/k_m$ from $x$ is to make the beam focused at the center of the domain.


Below we show the numerical results for the case (TBC1) and zero reflectivity. 
The numerical simulations for the case (TBC2) and $R(h)\neq 0$ give similar shape profiles. The inclusion of the reflectivity in the simulation makes the rupture time of the film  shorter. \textit{In this sense the film is more unstable when $R(h)\neq 0$.} We discuss this issue in more detail at the end of this Section.

\begin{center}
\textit{Impacts of the different modes of irradiation}
\end{center}

\begin{figure}[h!]
\centering
\includegraphics[width=3.8in,height=2.5in]{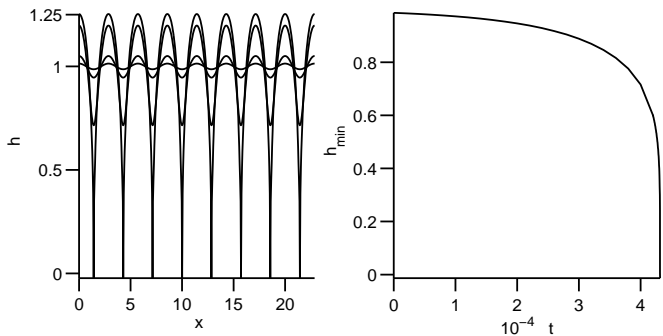}
	\caption{{\small{Profile of the film height (left), and the evolution of the minimum point on the film surface (right). The surface is heated by a uniform laser beam.}}}
	\label{SP_Uniform}
\end{figure}

\begin{figure}[h!]
\centering
	\includegraphics[width=3.4in,height=2.0in]{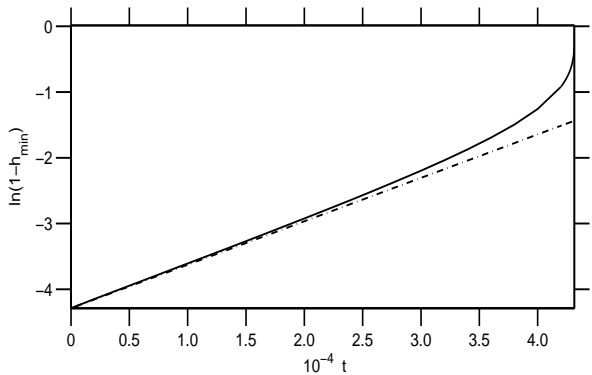}
	\caption{{\small{Plots of $ln(A_0)+\omega_{max}t$ (linear theory, dashed-dot curve) and $ln(1-h_{min}$) (nonlinear simulation, solid curve) vs time. The slope of the solid line equals the growth rate of the instability.}}}
	\label{Lin_NL}
\end{figure}

\emph{$f=1$.} Fig. \ref{SP_Uniform} shows the evolution of the film surface resulting from  uniform irradiation, both in time and space, by a single laser beam. The minimum point of the initial film surface goes further down and touches the substrate. In the simulation, the nonlinear dynamics of the film is followed until its
minimum height gets a very small value (0.001) and then the film is said to be ruptured. Favazza et al. \cite{FKS} estimate the \emph{rupture time} $T_r$ of a film (i.e., the time scale  of dewetting) as $\tilde T_r =2\pi/\omega(k_m)$. For a $10$nm-thick film their calculation gives $\tilde T_r \approx 1.25$ ms. Our linear stability analysis gives $\tilde T_r\approx 1.9$ ms. Note that $T_r$ depends on the amplitude of the initial surface perturbation. Naturally, perturbations with larger initial
amplitude rupture the film faster. In the nonlinear dynamical simulation shown in Fig. \ref{SP_Uniform} the
rupture time for $A_{0}=0.01375$ is approximately $0.9$ ms, which corresponds to the dimensionless time $T_r\approx 43150$.
It is seen in Fig. \ref{Lin_NL} that at the initial stage of the irradiation the nonlinear instability growth rate
matches the one predicted by the linear stability analysis, but towards the end of the simulation the nonlinear instability grows much faster. This can be explained by observing that when the film height gets small the factor $A/h$ in Eq. (\ref{E0Refl}) becomes very large, signaling that the destabilizing van der Waals component of the
disjoining pressure dominates over
stabilizing forces, which results in a faster instability growth.
We also noticed that smaller values of the capillary number, $C$, sometimes result in a ring rupture, meaning
that the film surface touches the substrate some distance from the vertical line through the minimum of the perturbation.
Such rupture leaves behind an array of small liquid drops \cite{AW}.

\emph{$f$ as in Eq. (\ref{sp}).}
The shape profile obtained by irradiating the surface with a single pulse of width $d=10787$ ($200 \mu s$  dimensional pulse width) is similar to the one resulting from uniform laser beam heating (Fig. \ref{SP_Uniform}). However,
$T_r \approx 35138$ in this case, which is less than the rupture time for the uniform heating. The reason is that the energy absorbed from the Gaussian beam is less than the energy absorbed from the uniform irradiation, which causes the rapid growth of the instability leading to fast rupture.

\emph{$f$ as in Eq. (\ref{np}).}

\begin{figure}[h!]
\centering
	\includegraphics[width=3.4in,height=2.5in]{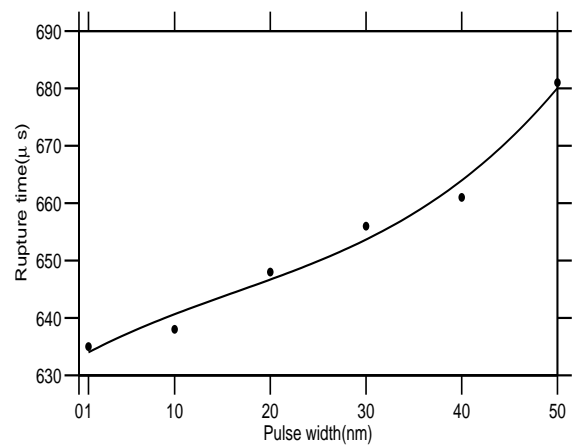} 
		\caption{{\small{Variation of the rupture time with the pulse width. (The curve is only the guide for the eye.)}}}
	\label{NP}
\end{figure}

In this simulation we use several sequences of Gaussian pulses. In each sequence, pulses have same width
and also the repetition frequency of pulses is same for all sequences.
Fig. \ref{NP}
shows the rupture time vs pulse width. The surface morphology is similar to the one obtained by uniform laser beam heating. It can be seen that the rupture time increases as the pulse width increases. Again, this is because the energy generated from a narrower pulse is smaller than the energy generated from a wider pulse, and this increases the growth rate of the instability. Heating the film surface with wider pulses also require more number of pulses for the film to rupture. For example, 3064 pulses are enough to get the film ruptured when irradiated with a 1ns pulse,  compared to 3293 pulses required for a 50ns pulse irradiation.
It must be noted here that in any irradiation mode the rupture
time increases approximately linearly with the peak intensity $I$. This effect can be seen from
Eqs. \rf{omegaRefl} and \rf{cutoffRefl}. Indeed, larger $I$ does not change the magnitude of the fourth (destabilizing) term there (see the definitions of the parameters $M,\ T_a$ and $T_s$ in the Table), but it increases the stabilizing contribution of the last term. The inclusion in the model of weak evaporation and the corresponding recoil pressure at the free surface  should partially reverse this trend \cite{AW}.

\begin{figure}[h!]
\centering
	\includegraphics[width=3.8in,height=3.4in]{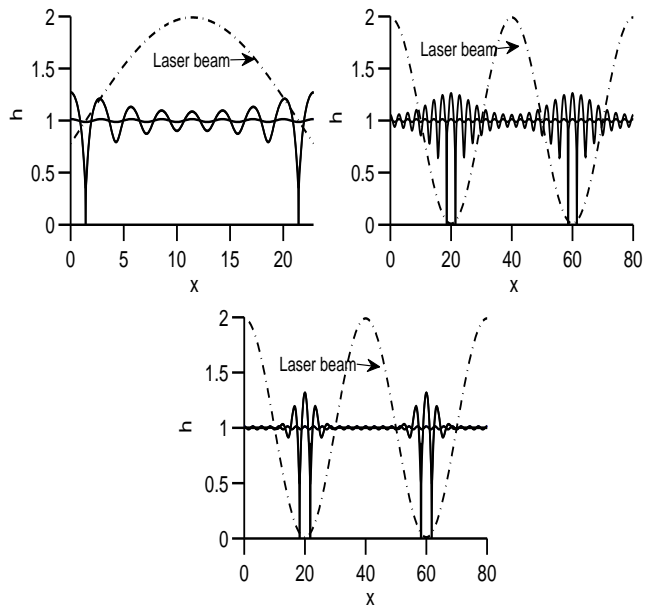}
	\caption{{\small{Profile of the film height after the two-beams interference heating. Top row, left: $H=10$ nm, the initial perturbed surface has eight wavelengths in the $x$-domain; Top row, right: $H=10$ nm, the initial perturbed surface has twenty eight wavelengths; Bottom row: $H=15$ nm, the initial perturbed surface has twenty eight wavelengths.}}}
	\label{twobeam}
\end{figure}

\emph{$f$ as in Eq. (\ref{fc}).}
Fig. \ref{twobeam} shows the film profile after the two-beams irradiation, together with the interference
field profile, i.e. the function $f(x)$. Values of the parameters are $\alpha=0.99,\ q=0.157$ and $k=2.2$.
The ruptures first occur in the low temperature regions,  while the strong heat generated in the high temperature regions makes the film surface stay near the equilibrium position $h=1$.
The rupture time is $600\mu s$, which is comparable to the time estimate given in Ref. \cite{TFTGKS} ($>100\mu s$  for a 10nm-thick film).
If the irradiation is
stopped immediately after the first ruptures occur, the solidification that follows is expected to create the regular array
of metallic ridges (nanowires)
with the axes along the $y$-direction. In contrast to Fig. \ref{SP_Uniform} where the irradiation is uniform,
here the ridges have different volume, with the large (small) volume in the cold (hot) regions. Thus the ridges volume distribution has the period $\ell$ of the interference imprint. This distribution is qualitatively consistent with the PLII experiments and modeling \cite{LTGZK}. Since the bulk film temperature is higher for thicker films (Fig. \ref{MaxT_bRefl}),
then increasing the film height from 10 nm to 15 nm eliminates most ridges, except the one in the destructive regions
(the bottom figure in Fig. \ref{twobeam}).
(Note that scales are very different along the $x$ and $h$ axes in Figs. \ref{SP_Uniform} and \ref{twobeam}, thus the true
cross-section of each ridge is closer to the circular than it appears.)
It must be noted that by nature of this model the substrate is exposed only at the points of film rupture between the ridges in the cold regions, while in the experiment each ridge terminates at the substrate. In other words, the simulation with this model can't be continued for $t > T_r$ and thus we can't predict how the mass will be re-distributed if the irradiation persists.\\

\begin{figure}[!h]
\centering
	\includegraphics[width=3.4in,height=2.2in]{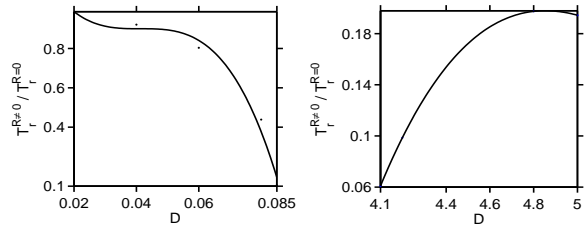}
	\caption{{\small{Variation of the ratio of the rupture times for the simulations with and without reflectivity, with $D$. (The curve in the left panel is only the guide for the eye.)}}}
	\label{CompTr_R}
\end{figure}

Finally, in Fig. \ref{CompTr_R} we plot the ratio of the rupture times, $T_r^{R\neq 0}/T_r^{R=0}$ vs. $D$ for the case (TBC2).
The simulations were done, in each case, with the most dangerous wavenumber $k_{m}$ (for $R=0$ and $R\neq 0$ the
most dangerous wavenumbers are different). 
It can be seen that this ratio is less than one for all values of $D$ and changes non-monotonously with $D$, decreasing first
and then increasing. The minimum value is as small as $\sim 10^{-2}$.
The ratio is less than one because
the reflectivity 
reduces the  heat generation in the film (Eq. \rf{Source}), thus reducing the stabilization. 
Note that, since the simulation is started with the small surface deformation
(such that the predictions of the linear stability theory
are valid for at least some time), the ratio does not give meaningful
comparison in the interval $D\sim 1$. This is because the surface is linearly stable for $R=0$ and $R\neq 0$ when
$D\sim 1$, as shown in Fig. \ref{omaxRefl}, and both $T_r^{R\neq 0}$ and $T_r^{R=
0}$ are formally infinite. Thus the ratio is not plotted for $0.085\le D\le 4.1$ 

\section{Discussion and Conclusions}

\label{Sec5}

In this paper we study the dewetting dynamics of a pulsed laser-irradiated  metallic films. A lubrication-type model describing the flow of a molten film and the heat conduction in the film is developed. The heat absorbed from the laser beam is included as a source term in the heat conduction problem and the temperature field distribution in the film is obtained by solving this problem analytically. In the laser interference mode of irradiation, we observe that the lateral temperature distribution in the film mimics the shape of the lateral power intensity distribution of the laser.
The temperature difference across the film and the temperature in the film are higher for optically thick films than
for the optically thin ones.

The temperature field is used to derive the 3D long-wave evolution PDE for the film height. In order to get clear understandings of the film dynamics, we study  the 2D version of the equation by means of the linear stability analysis and  numerical simulations.

The linearized problem allows us to investigate the stabilizing and destabilizing effects of various system parameters.
Higher
peak intensity of the beam and larger Marangoni number $M$
either delay the
rupture time of an initially perturbed film or make the perturbation decay, while smaller surface Biot number $\beta$ and
substrate Biot number $\beta_s$ have the same effect. The increasing optical thickness $D$ can have either 
stabilizing or destabilizing effect, depending on the magnitudes of the film reflectivity and the ratio 
of the substrate to film thermal conductivities. As film becomes thinner, the stabilizing effect of the internal 
heat generation becomes smaller as more heat is generated in thicker films. 

Impacts of the different modes of irradiation are investigated numerically in the 2D setting. For the spatially uniform (single beam) irradiation the film rupture is spatially periodic with the wavelength of the fastest growing perturbation. The latter wavelength is determined by values of all system parameters, including the laser parameters. In the two-beam interference heating mode the ruptures occur in the (cold) regions of destructive interference, while at the (hot) regions of constructive interference the initial surface perturbation is still developing. Assuming the irradiation is stopped after the first ruptures, the solidification is expected to create a ridge (nanowire) array, where the spatial distribution of ridges volumes follows the spatial periodicity of the interference imprint. 

These conclusions do not depend, at large, on the presence of the thickness-dependent reflectivity.
However, quite unexpectedly we found that reflective films with $D\sim 1$ can be completely stabilized against
dewetting and rupture, although films with  $D$ either small or large are less stable than the corresponding non-reflective films (due to smaller magnitudes of the heat source in the reflective films).
The rupture time from the simulations is comparable to the estimated and the experimentally obtained values \cite{FTKS}. 
The rupture time of the films having nonzero reflectivity is significantly shorter than the one of the non-reflective films.

Finally, we point out the difference of thermocapillary mechanisms with and without
heat generation in the film due to laser beam irradiation of the film surface.
In standard applications, when the substrate is heated up (i.e., the situation where the film surface is not irradiated and thus there is no  heat production in the film)
the thermocapillary effect is governed
by the $M\beta (T_a-T_s)h^2\nabla h$ term in the evolution equation, where $T_a < T_s$.  This term is responsible for
the fluid flow from the high temperature region (the one that is closer to the hot substrate, i.e.
the trough of the surface undulation) to the low temperature region, i.e. the crest of the surface undulation,
resulting in instability and ultimate film break-up in the high $T$ region.
In that case the temperature gradient across the film is negative, $\partial T/\partial z < 0$.
On the other hand, when the film surface is irradiated and the heat is generated
in the film, the temperature gradient $\partial T/\partial z > 0$ (see Fig. \ref{Temp}), despite that
still $T_a < T_s$.
The 
heat source term in the evolution equation
counterbalances the standard term, reversing the direction of the fluid flow. 
In the multiwavelength nonlinear simulation shown in Fig. \ref{twobeam} this effect 
manifests as enhanced surface stability in the hot regions.
The linear stability is also drastically affected (see Figs. \ref{omegaVk}, \ref{KcVM}, \ref{KcVdH} and \ref{omaxRefl}).
These results could help understanding the dewetting and rupture process in ultrathin metal films irradiated by pulsed 
laser beams, such as Co, Fe, Au, Ni, Cu, Ag and Mo films on glass and SiO$_2$/Si substrates 
\cite{BSHL,HCS,FKS,TTFSK,TFTGKS,LTGZK,KVR,FTKS,FTKKS}.

Future work will focus on development of accurate and efficient numerical methods (finite difference and spectral)
for the 3D evolution equation, simulations of the corresponding film dynamics, and on quantitative characterization of 
the 3D structures size and ordering.



\vspace*{0.5cm}

\section*{Acknowledgements}
MK acknowledges Ramki Kalyanaraman for several useful discussions, and Sergey Shklyaev for commenting on the first
draft of the paper. Anonymous Referee is acknowledged for many useful suggestions.


\begin{thebibliography}{200}

\bibitem{BSHL} J. Bischof, D. Scherer, S. Herminghaus, and P. Leiderer,
\textit{Phys. Rev. Lett.}$\;${\bf 77}, 1536 (1996).

\bibitem{V} A. Vrij,
\textit{Discuss. Faraday Soc.}$\;$ {\bf 42}, 23 (1966).

\bibitem{RJ} E. Ruckenstein and R.K. Jain,
\textit{Faraday Trans. 2}$\;$ {\bf 70}, 132 (1974).

\bibitem{BWD} F. Brochard-Wyart and J. Daillant,
\textit{Can. J. Phys.}$\;${\bf 68}, 1084 (1990).

\bibitem{GKS}
A.~Ghatak, R.~Khanna, and A.~Sharma, \textit{J. Colloid Interface Sci.}$\;$
{\bf 212}, 483 (1999).

\bibitem{SHJ}
R.~Seemann, S.~Herminghaus, and K.~Jacobs, \textit{J. Phys.: Condensed
Matter}$\;$ {\bf 13}, 4925 (2001).


\bibitem{HCS} S.J. Henley, J.D. Carey, and S.R.P. Silva,
\textit{Phys. Rev. B}$\;$ {\bf 72}, 195408 (2005).


\bibitem{FKS} C. Favazza, R. Kalyanaraman, and  R. Sureshkumar,
\textit{J. Appl. Phys.}$\;$ {\bf 102}, 104308 (2007).

\bibitem{TTFSK} J. Trice, D. Thomas, C. Favazza, R. R. Sureshkumar, and R. Kalyanaraman,
\textit{Phys. Rev. B}$\;${\bf 75}, 235439 (2007).

\bibitem{TFTGKS} J. Trice, C. Favazza, D. Thomas, H. Garcia, R. Kalyanaraman, and R. R. Sureshkumar, 
\textit{Phys. Rev. Lett.}$\;${\bf 101}, 017802 (2008).

\bibitem{LTGZK} L. Longstreth-Spoor, J. Trice, H. Garcia, C. Zhang, and R. Kalyanaraman,
\textit{J. Phys. D: Appl. Phys.}$\;$ {\bf 39}, 5149 (2006).

\bibitem{KVR} Yu. Kaganovskii, H. Vladomirsky, and M. Rosenbluh,
\textit{J. Appl. Phys.}$\;$ {\bf 100}, 044317 (2006).

\bibitem{FTKS} C. Favazza, J. Trice, R. Kalyanaraman, and R. Sureshkumar,
\textit{Appl. Phys. Lett.}$\;${\bf 91}, 043105 (2007).

\bibitem{FTKKS} C. Favazza, J. Trice, H. Krishna, R. Kalyanaraman, and R. Sureshkumar,
\textit{Appl. Phys. Lett.}$\;$ {\bf 88}, 153118 (2006).

\bibitem{WD} M.B. Williams and S.H. Davis,
\textit{J. Colloid Interface Sci.}$\;${\bf 90}, 220 (1982).

\bibitem{AW} V.S. Ajaev and D.A. Willis,
\textit{Phys. Fluids}$\;${\bf 15}, 3144 (2003);
\textit{Numer. Heat Transfer, Part A}$\;${\bf 50}, 301 (2006).

\bibitem{BG} A.S. Basu and Y.B. Gianchandani,
\textit{Appl. Phys. Lett}$\;${\bf 90},  034102 (2007).

\bibitem{H} F.J. Higuera,
\textit{Phys. Fluids}$\;${\bf 12}, 2186 (2000).

\bibitem{OP} A. Oron and Y. Peles,
\textit{Phys. Fluids}$\;${\bf 10}, 537 (1998).

\bibitem{O1} A. Oron,
\textit{Phys. Fluids}$\;${\bf 12}, 29 (2000).

\bibitem{G} R.O. Grigoriev,
\textit{Phys. Fluids}$\;${\bf 14}, 1895 (2002).

\bibitem{Boyd} I.W. Boyd, Laser Processing of Thin Films and Microstructures,
\textit{Springer-Verlag}$\;$ (1987) 320 p.

\bibitem{Derjaguin} B.V. Derjaguin, L.F. Leonov, and V.I. Roldughin,
\textit{J. Colloid Interface Sci.}$\;$ {\bf 108}, 207 (1985); also in: \textit{Prog. Surf. Sci.}$\;$ {\bf 40},
232 (1992).

\bibitem{ODB} A. Oron, S.H. Davis,  and S.G. Bankoff,
\textit{Rev. Mod. Phys.}$\;$ {\bf 69}, 931 (1997).

\bibitem{ZB} L. Zhornitskaya  and A.L. Bertozzi,
\textit{SIAM J. Numer. Anal.}$\;$ {\bf 37}, 523 (2000).


\end{thebibliography}
\end{document}